# Spin accumulation in forward-biased MnAs/GaAs Schottky diodes

J. Stephens[1], J. Berezovsky[1], J. P. McGuire[2], L. J. Sham[2], A. C. Gossard[1], and D. D. Awschalom[1]

[1]*Center for Spintronics and Quantum Computation, University of California, Santa Barbara, CA 93106*

[2] *Department of Physics, University of California San Diego, La Jolla CA 92093-0319*

## Abstract

We describe a new means for electrically creating spin polarization in semiconductors. In contrast to spin injection of electrons by tunneling through a reverse-biased Schottky barrier, we observe spin accumulation at the metal/semiconductor interface of forward-biased ferromagnetic Schottky diodes, which is consistent with a theory of spin-dependent reflection off the interface. Spatiotemporal Kerr microscopy is used to image the electron spin and the resulting dynamic nuclear polarization that arises from the non-equilibrium carrier polarization.

PACS numbers: 76.70.Fz, 78.47.+p, 75.70.-i, 72.25.Dc

The injection of electrons from ferromagnetic metals into semiconductors has recently received much attention in the field of spintronics since these systems have the potential to serve as room-temperature sources of spin polarization[1]. To date, most research in this vein has focused on electron currents flowing through a tunnel barrier from the ferromagnet (FM) to the semiconductor (SC). For example, spin injection has been observed for tunneling through Schottky[2,3] and aluminum oxide[4] tunneling barriers as well as in more complicated structures such as magnetic tunnel transistors[5]. All of these schemes share the common feature that spin-polarized electrons are injected from the FM into the SC. However the reverse process, in which electrons flow from SC to FM is also spin-dependent[6,7]. In addition, the recent discovery of the ferromagnetic proximity polarization effect[8,9] demonstrated that initially unpolarized optically excited electrons become rapidly spin polarized at a forward-biased FM/SC interface.

Here we report evidence of spin accumulation in ferromagnetic MnAs/GaAs Schottky diode structures biased in the forward direction with no optical excitation. As an electron current flows from the GaAs into the MnAs, a net electron spin polarization arises in the GaAs. We use two spatially resolved techniques to map the lateral dependence of the electron spin polarization and the subsequent dynamically polarized nuclear spins. Both the electron and nuclear spin polarization profiles track the expected current density through the Schottky diode. A theory in which the spin polarization arises due to spin-dependent electron transmission and reflection at the FM/SC interface is used to estimate that the steady state spin polarization at modest forward bias in our device is on the order of a few percent.[10,11]

Samples are grown by molecular beam epitaxy and consist of 25 nm type-A MnAs, 500 nm n-GaAs, and 400 nm $Al_{0.75}Ga_{0.25}As$, grown on a semi-insulating GaAs(100) substrate[12]. Using photolithography and wet etching, a 50 x 100 μm MnAs mesa is defined at the center of a channel etched into the n-GaAs layer, as shown in Fig. 1a. A metal contact is deposited on the n-GaAs across the channel from the MnAs mesa, and annealed to form an ohmic contact. A second contact is then deposited on the MnAs mesa, thus forming a Schottky diode in series with a resistor, which shows typical rectifying I-V characteristics. This lateral geometry is essential for many proposed spintronic devices. The sample is epoxied to sapphire and a chemically-selective wet etch is used to form a window in the substrate, allowing optical access to the sample from both sides. Hysteresis loops measured in a superconducting quantum interference device (SQUID) magnetometer demonstrate that the MnAs is ferromagnetic with a Curie temperature of ~320 C. While the data presented in this paper are from a single sample with the n-GaAs layer Si-doped at 8e16 $cm^{-3}$, we have measured a number of samples with doping ranging from 6e16 to 1e17 $cm^{-3}$ and obtained qualitatively similar results. We have also fabricated a control sample with a non-magnetic Schottky diode in which none of the spin effects discussed here were present.

In order to probe electron and nuclear spin polarization in the GaAs epilayer, we use two different optical techniques. In each case the sample is mounted in a He flow cryostat, and laser pulses which are used to probe the spin polarization (see below) are focused on the sample through a scanning microscope objective, providing micron-scale lateral resolution.[13] Unless otherwise specified, all of the studies were performed at T = 7.5 K measured with a thermocouple mounted several millimeters from the sample.

To measure the electron spin, we use a hybrid of the Hanle and magneto-optic Kerr effect (MOKE) measurements, which we refer to as Hanle-MOKE. In this technique, the polarization of a linearly polarized probe beam ($\lambda \sim 812$ nm) is rotated upon reflection off of the sample through an angle proportional to the projection along the growth direction of the time-averaged electron spin polarization. This angle is measured as an applied transverse magnetic field is swept in a small range about $B_{app} = 0$. This contrasts with traditional Hanle experiments, in which the spin is detected by photoluminescence polarization. For electron spin initially polarized at $t = 0$ perpendicular to the applied field and in the plane of the sample, the MOKE signal may be written as $S(t) = A\exp(-t/\tau)\sin(\omega_L t)$, where $\tau$ is the transverse spin lifetime, $\omega_L = 2\pi\nu_L$ where $\nu_L$ is the Larmor precession frequency, and A is a scale factor. Integrating S with respect to t, one finds the time averaged signal to be $<S> = A\omega_L/(1/\tau^2 + \omega_L^2)$ (Eq. 1). Fitting the data obtained by this technique yields the spin lifetime and the parameter A, which is proportional to the spin polarization rate. The quantity $A\tau$ is then proportional to the time averaged spin polarization in zero applied field, assuming the spin lifetime is independent of field (a good assumption for the modest fields used here).

Non-equilibrium electron spin polarization should result in the well-known phenomenon of dynamic nuclear polarization (DNP), which is due to electron and nuclear spin-flip interactions[14]. Moreover, DNP was previously observed to arise from spin injection in reverse-biased Schottky diodes[15]. To measure the nuclear polarization we use the time-resolved Kerr rotation (TRKR) technique[16]. Here, a ~150 fs duration circularly-polarized pump pulse ($\lambda \sim 812$ nm) from a mode-locked Ti:Sapphire laser excites spin-polarized electrons into the conduction band of the n-GaAs. These optically injected spins then precess around a transverse magnetic field at the Larmor frequency and decay exponentially with transverse lifetime $\tau$. A time $\Delta t$ later, a linearly-polarized probe pulse measures the instantaneous projection of the spin onto the growth direction through the Kerr effect. Scanning $\Delta t$ yields an oscillatory Kerr rotation signal $\theta_K(\Delta t) \sim e^{-(\Delta t/\tau)} \cos(2\pi\nu_L\Delta t)$ from which one obtains $\nu_L$. Through the hyperfine interaction, a net nuclear polarization acts like an effective magnetic field, $B_N$, on the electron spins; therefore, the Larmor frequency of the optically injected electrons is proportional to the sum of the applied field $B_{app}$ and $B_N$.[17] By measuring $\nu_L$, we obtain the effective nuclear field which is proportional to the nuclear polarization[18].

Characteristic Hanle-MOKE and TRKR curves, along with schematics of each measurement geometry are shown in Figs. 1b and 1c.[19] For the Hanle-MOKE measurement, the sample is mounted with the magnetocrystalline easy axis of the MnAs perpendicular to the applied field. Due to the strong uniaxial anisotropy present in MnAs, the magnetization will deviate only slightly from the easy axis in the magnetic fields used in these experiments. One effect of this is that in a 2 kG external field, the dynamic nuclear polarization is an order of magnitude smaller than in the parallel geometry, and at lower fields it is negligible. We apply a 6 kHz square-wave to the Schottky diode, oscillating between $V_b = 0$ and $+1.5$ V ($I_b = 0$ to $1.65$ mA), which allows us to use lock-in detection of the electrically-induced MOKE signal. Fitting the curves obtained through the Hanle-MOKE technique yields the spatial map of the time-averaged spin polarization ($A\tau$) shown in Fig. 2a. As expected, we find that switching the magnetization of the FM yields a sign change in the Hanle-MOKE signal (not shown).

For the nuclear polarization measurements, the sample was mounted with the MnAs easy axis parallel to the applied magnetic field. In an applied field of $B_{app}$ = 2 kG and a forward bias of $V_b$ = 1.5 V ($I_b$ = 1.65 mA), we observe nuclear polarization under the MnAs mesa building up over the course of approximately 20 minutes, with a maximum effective nuclear field $B_N$ = 0.45 T and a maximum effective field gradient of $8 \times 10^4$ T/m. The presence of nuclear polarization implies that a non-equilibrium electron spin polarization is being maintained in the SC, leading to DNP. The sign of the effective nuclear field indicates that the electron spin is polarized anti-parallel to the MnAs magnetization, in agreement with previous measurements.[9] Fig. 2b shows an image of the resulting nuclear polarization which is in qualitative agreement with the electron polarization. Discrepancies between the images of electron and nuclear polarization may be due in part to the fact that electron spins decay with the transverse lifetime in the case of the Hanle-MOKE measurement but the longitudinal lifetime is the relevant time-scale in the DNP measurement. Sweeping the field and measuring the nuclear polarization yields a hysteresis loop similar to those previously measured[8], with a switching event occurring at the MnAs coercive field, thus confirming the role of the FM in this phenomenon.

We have used both the TRKR and Hanle-MOKE measurements to investigate the dependence of nuclear and electron spin polarizations on the forward-bias current. (The tunnel barrier is too large to see any reverse-bias spin injection.) Fig. 2c shows linecuts of the electron spin polarization taken along the path indicated by the dashed line in the inset of Fig 1a. We find no measurable electron polarization for $V_b$ < 0.8 V, which corresponds to the turn-on of the Schottky diode. At higher voltages, current begins flowing and the electron polarization increases. The maxima of these curves as a function of applied bias are shown in the inset to Fig. 2c.

The same line-cuts were measured using the TRKR technique, and the extracted nuclear spin polarization is shown in Fig 2d. The results closely match the electron polarization profiles. Again, the maxima of these curves are shown in the inset. One difference between the two measured spin profiles is that the nuclear polarization reaches a maximum at about $V_b = 1.6$ V and subsequently decreases at higher bias, possibly due to resistive heating.

We also performed temperature dependence measurements of both the nuclear and electron spin, the results of which are shown in Fig 3. A TRKR measurement with optically injected spins at zero applied field was used to normalize the Hanle-MOKE results as a function of temperature since the magnitude of the Kerr effect was observed to decrease by ~30% from T = 7.5 K to T = 50 K. We find that the magnitude of the nuclear polarization and the time-average electron spin polarization both decrease monotonically with increasing temperature, but that the nuclear polarization decreases more quickly. This is expected since the nuclear polarization depends not only on the average electron spin polarization but also on the nuclear relaxation time, which decreases with increasing temperature[20].

Figures 4a and 4b show the same current dependence measurements as Fig 2d, but with the spin accumulation rate A and spin depolarization rate $1/\tau$ plotted separately. We see that the spin accumulation rate is sharply peaked near the leading edge of the MnAs, and is quite small elsewhere under the mesa. Similarly, the spin depolarization rate is peaked at the leading edge, but then decreases to a constant value farther away. The shape of A as a function of position is similar to what one would expect for the lateral current density across the SC/FM interface, since the conductivity of the GaAs is much lower than that of the metallic MnAs. The strongly enhanced spin depolarization rate in the vicinity of the MnAs leading edge is likely due to transport of spin into the FM or away from the probe spot within the SC, an effect which will be greatest near the front of the magnetic mesa where the current density is largest. The temperature dependence of A and $1/\tau$ are shown in Fig 4c and 4d (from the measurements of Fig. 3b). Here, the spin lifetime appears to be dominated by a temperature independent contribution near the edge of the MnAs mesa, which is again consistent with the picture of spin being transported away from the probe spot in the area of high current density.

The observed effect may be qualitatively described by a simple model of spin-dependent reflection off the FM/SC interface.[10,11] The ferromagnet is modeled as exchange-split parabolic bands, resulting in different wavevectors in the ferromagnet for the two spin channels at a given energy.[21] The wavefunction matching conditions at the interface result in spin-dependent reflection and transmission coefficients for electrons incident at the interface. We consider transport across the interface under forward-bias (see schematic shown in Fig. 1d), ignoring the spatial dependence of the current density and assuming the spin-flip time is sufficiently long such that the two spin channels are independent. Using the Drude relation, the spin polarization due to reflection can be calculated from the spin-dependent currents flowing through the interface,

$$\frac{n_\uparrow - n_\downarrow}{n_\uparrow + n_\downarrow} = \frac{\int_0^{k_f} dk_z k_z \sqrt{k_f^2 - k_z^2}\left[T_\uparrow - T_\downarrow\right]}{2\pi^2 n + \int_0^{k_f} dk_z k_z \sqrt{k_f^2 - k_z^2}\left[T_\uparrow + T_\downarrow\right]},$$ where $k_f$ is the Fermi wavevector in

the semiconductor, $k_z$ is its component in the growth direction, $n$ is the doping density,

$T_\uparrow$ and $T_\downarrow$ are the transmission coefficients for the two spin channels.  Using reasonable material parameters[22] and assuming the SC spin lifetime is dominated by current flow into the FM, the theory estimates the spin polarization to be a few percent and is consistent with the observed nuclear polarization.[15]

The spin accumulation effect represents a new method for electrically generating spin polarizations in semiconductors.  Instead of injecting electrons from a ferromagnet into a semiconductor, we find that spin polarized electrons accumulate at the SC/FM interface under forward bias.  This is consistent with a theory of spin-dependent reflection, which raises the interesting possibility of electrical spin injection without transport across an interface.  This phenomenon may find application in future spin-based devices; in addition, the spatially-resolved measurement techniques used here should prove useful in the characterization of such structures. Work supported by the ARO and DARPA/ONR.

**Figure Captions**

Fig. 1. (a) Charge-coupled device micrographs of spin accumulation device. Closeup depicts the MnAs magnetic easy axis direction as well as the linecut measurement path (see text). (b) TRKR and (c) Hanle-MOKE data (symbols) and fits (lines). Insets show the two measurement geometries. (d) Schematic conduction band diagram of Schottky interface for magnetization parallel to spin-up electrons. Also depicted is a cartoon of the spin-dependent incident ($I_I$), reflected ($I_R$), and transmitted ($I_T$) electron currents.

Fig. 2. (a) 2D image of effective nuclear field $B_N$ taken at T = 7.5 K with forward bias of 1.5 V and applied field of 2 kG. (b) 2D image of time averaged electron spin ($A\tau$) taken at T = 7.5 K with a square-wave between 0 V and 1.5 V applied at 6 kHz. Dotted lines show outline of MnAs mesa. (c) Linecuts along the path indicated in Fig. 1a of $B_N$ and (d) $A\tau$ for various forward biases between 0.8 and 2.2 V. Insets show the peak value of each as a function of bias.

Fig. 3. Linecuts at various temperatures taken along the path shown in Fig. 1a of (a) effective nuclear field and (b) time-averaged electron spin from T = 7.5 K to T = 50 K. Dashed vertical line indicates the position of the front edge of the MnAs mesa. For the nuclear field measurement the forward bias was 1.5 V while for the electron spin measurement a 6 kHz square wave between 0 V and 1.5 V was applied.

Fig. 4. Linecuts showing the bias dependence of (a) the electron spin accumulation rate A and (b) the electron spin depolarization rate $1/\tau$ taken at T = 7.5 K. Linecuts of A and $1/\tau$ for T = 7.5 K to T = 50 K are shown in (c) and (d), respectively. The bias conditions are the same as in Fig. 3.

This could be an artifact due the Hall effect, or to a small deviation of the initial direction of the spin from the sample plane. Nevertheless, we use Eq. 1 to fit the measured curves, effectively averaging out the asymmetry.

[20] A. Abragam, Principles of Nuclear Magnetism (Oxford University Press, Oxford, 1961).

[21] This simplified model is meant to mimic the spin-dependence in the ferromagnet; the spin-dependence in the reflection and transmission coefficients and the magnitude of the polarization should be similar to what the simple model predicts if the realistic band structure were used, although the sign of the polarization would depend on the spin-dependent wavevectors at the appropriate energy for transmission.

[22] GaAs: effective mass $m^* = 0.07$, mobility $\mu = 3000$ cm$^2$/V-s, doping n = 8e16 cm$^{-3}$, MnAs: $m^* = 1$, spin-up potential $U_{fm}^{\uparrow} = 3.0$ eV, spin-down potential $U_{fm}^{\downarrow} = 1.5$ eV, Schottky barrier height: $U_b = 0.8$ eV.

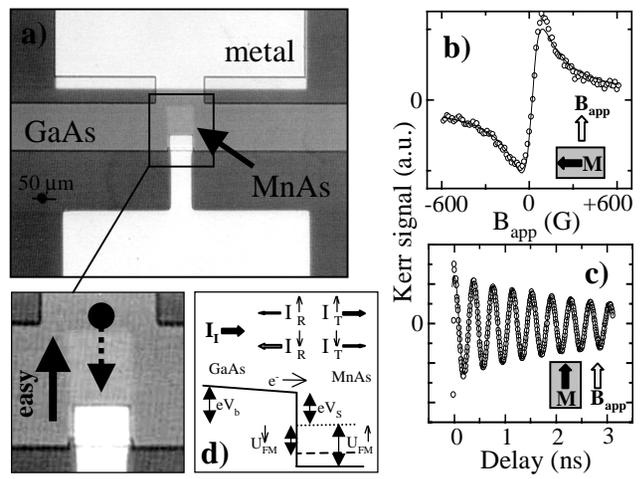

Figure 1

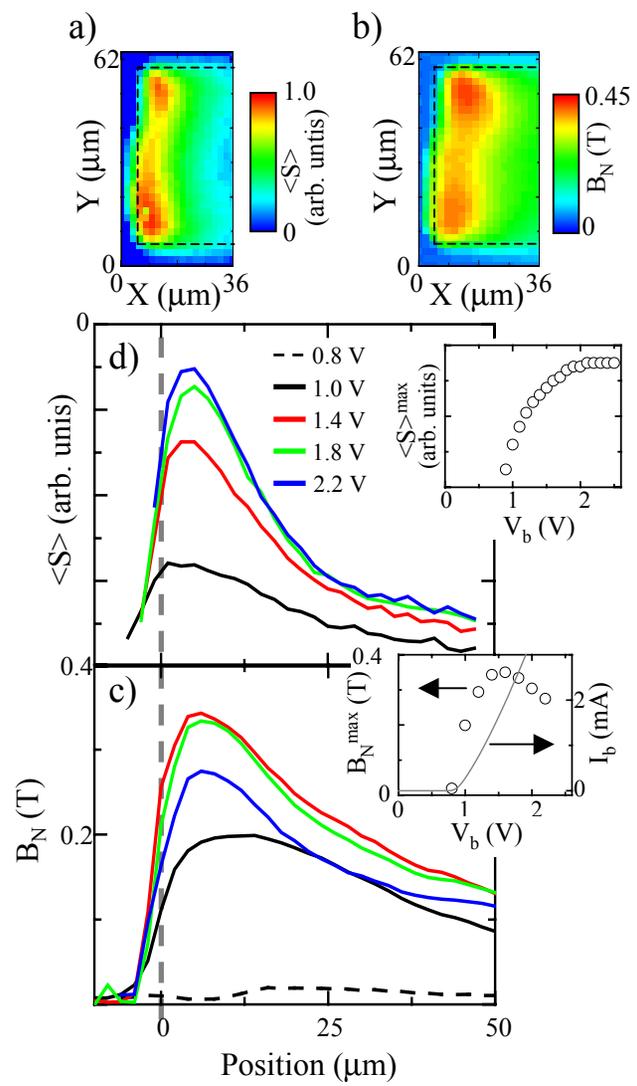

Figure 2

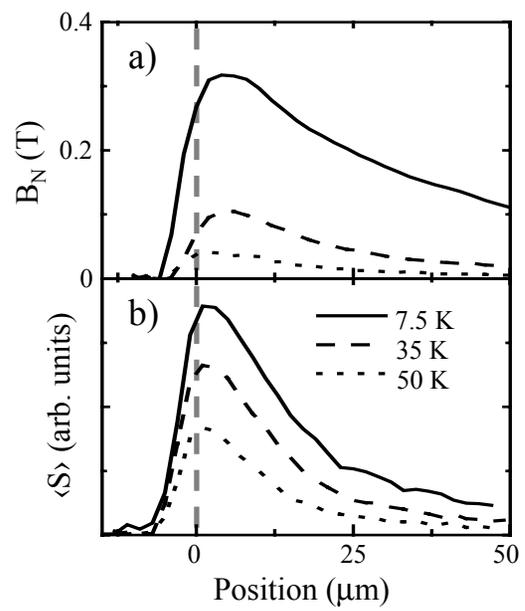

Figure 3

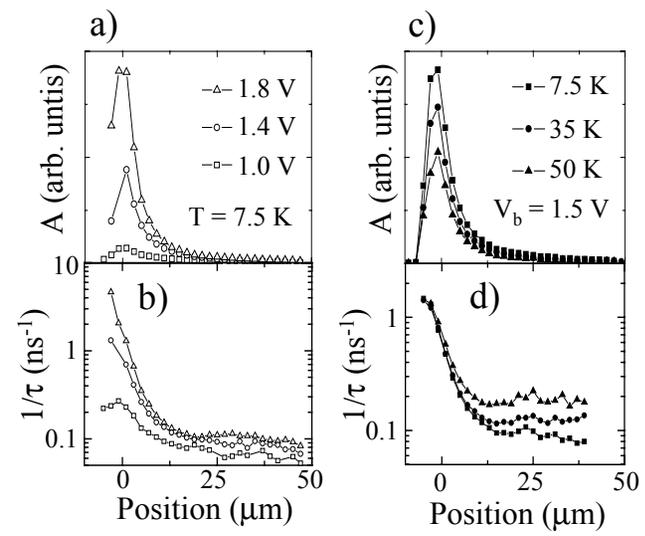

Figure 4